\documentclass[11pt,a4paper]{article}
\usepackage{epsfig}
\usepackage{amsmath}
\usepackage{amssymb}
\usepackage{verbatim}
\usepackage{bm}
\usepackage{color}
\usepackage{ulem}
\usepackage{cite}

\begin{document}

\newcommand{\solarprop}{\texttt{SOLARPROP}}
\newcommand{\galprop}{\texttt{GALPROP}}
\newcommand{\dragon}{\texttt{DRAGON}}
\newenvironment{code}{
\setlength{\parskip}{2mm}
\setlength{\parindent}{0pt}
\ttfamily\par
}
{\\[2mm]}

\thispagestyle{empty}
\begin{center}

{\LARGE\bf SOLARPROP: Charge-sign Dependent Solar Modulation for Everyone
}\\[12mm]

{\large Rolf Kappl\footnote{kappl@th.physik.uni-bonn.de}} 
\\[6mm]
{\it Bethe Center for Theoretical Physics\\
and\\
Physikalisches Institut der Universit\"at Bonn\\
Nussallee 12, 53115 Bonn, Germany
}
\vspace*{12mm}
\begin{abstract}
We present \solarprop{}, a tool to compute the influence of
charge-sign dependent solar modulation for cosmic ray
spectra. \solarprop{} is able to use the output of popular tools like
\galprop{} or \dragon{} and offers the possibility to 
embed new models for solar modulation.  
We present some examples for proton, antiproton and positron fluxes in the light
of the recent PAMELA and AMS-02 data.
\end{abstract}
\end{center}
\clearpage

\section{Introduction}

The study of high energetic cosmic rays is a well established method
to constrain possible decay or annihilation of 
dark matter to standard model particles. Cosmic rays may either have
secondary origin e.g. from spallation of primary particles or
originate from primary sources like supernova remnants, pulsars or
the decay or annihilation of dark matter. The
additional cosmic ray flux from dark matter would increase the
measured experimental values at earth. To determine the astrophysical background for
cosmic rays as well as their possible fraction originating from dark
matter, particles have to be propagated through the galaxy from
their origin to the solar system. The propagation of charged cosmic rays is well studied
in the literature~\cite{Maurin:2001sj,galprop,Strong:1998pw,dragon,Evoli:2008dv} and despite the still large uncertainty, an enormous
progress has been made during the last years. As an outcome
freely available tools like \galprop{}~\cite{galprop,Strong:1998pw} and \dragon{}~\cite{dragon,Evoli:2008dv} exist,
which simulate the propagation of cosmic rays through the
galaxy. 

Once, the cosmic rays have reached the solar system, they are
further influenced by the solar wind and the magnetic field of the
sun~\cite{Parker19659}. This effect is called solar modulation. It can be approximately taken into account with the so-called
force-field method~\cite{Gleeson:1968zza}. This approximation is independent of the charge of the cosmic
ray particles. The strength of solar modulation depends on the solar activity and thus on the 22 year solar cycle. High solar activity results in a stronger effect and thus a lower cosmic ray flux, whereas lower solar activity results in a smaller effect and a higher cosmic ray flux at the top of the earth's atmosphere.
The study of solar
modulation is a field of intensive research and many models which are
able to provide a better estimate than the force-field approximation for the influence of the heliosphere
on cosmic ray transport have been proposed (see~\cite{gast,Cholis:2015gna} for recent effective approaches based on the force-field approximation). 

Solar modulation can be well described by a Fokker-Planck equation
~\cite{Parker19659}. It has become popular to rewrite this equation to a set of
stochastic differential equations (see e.g.~\cite{yamada,zhang,Bobik:2011ig,strauss,pei,Alanko-Huotari,Potgieter:2013cwj}) which can be solved with Monte Carlo methods. \solarprop{}~\cite{solarprop} is a simple
Monte Carlo tool which is flexible enough that arbitrary models for
solar modulation in the heliosphere can be embedded. It
includes several simple models which show the effect of a more
realistic model for solar modulation. We outline that charge-sign
dependent models nicely fit the experimental data and are an advantage
to the force-field approximation. 
The force-field approximation is based on one free parameter, the so-called Fisk potential \(\phi\). It serves as a free fit parameter, adjusted a posteriori to model the effect of solar modulation. In contrast, the models in \solarprop{} are based on experimental values of the solar magnetic field. This makes it possible to model solar modulation without a fit parameter but in a predictive framework. For any given date in the past, different experimental values for the solar activity like the tilt angle of the heliospheric current sheet are used to model the effect of solar modulation in a predictive way.      

A better understanding of solar modulation is especially interesting for
antiprotons as their flux peaks around a kinetic energy of a few GeV, an
energy range where the effect of solar modulation is very
strong. Due to the low astrophysical background, antiprotons have been used to constrain dark matter models (see e.g.~\cite{Kappl:2011jw} for an analysis of the BESS-Polar II data). A lot of balloon based data are available at a few GeV and a proper treatment of solar modulation can help to decrease the uncertainties on indirect dark matter detection.

\solarprop{} is easy to use and easy to extend with new
custom models. The program is able to deal with input in FITS format
from \galprop{} and \dragon{}, as well as with input from
text files. The output is provided in the same formats. This makes
it possible to include \solarprop{} in the chain of cosmic ray
propagation tools. A realistic treatment of
cosmic ray propagation from the production in the galaxy to the detection at the top
of the earth's atmosphere is thus possible.  

\section{Physical background}

The transport equation for cosmic rays in the heliosphere is given by
~\cite{Parker19659, parker}
\begin{equation}
\frac{\partial f}{\partial t} = \nabla\cdot (\kappa\cdot \nabla f) -
(\mathbf{V}+\mathbf{V}_D)\cdot \nabla f+\frac{1}{3}(\nabla\cdot
\mathbf{V}) \frac{\partial f}{\partial\ln p}.
\end{equation}
\(f\) is the particle phase space distribution function,
\(\kappa\) the diffusion tensor, \(\mathbf{V}\) the solar wind
velocity, \(\mathbf{V}_D\) the particle drift velocity in the
heliospheric magnetic field and \(p\) the particle momentum. This
Fokker-Planck equation can be rewritten to an equivalent set of stochastic
differential equations (SDEs)~\cite{Gardiner1989,kopp}. The idea has been used by many
authors~\cite{yamada,zhang,Bobik:2011ig,strauss,pei,Alanko-Huotari,Potgieter:2013cwj} to solve the propagation of cosmic rays in the 
heliosphere with Monte Carlo techniques.

The SDEs can be written as
\begin{equation}
dx_i=A_i(x_i)dt+\sum_jB_{ij}(x_i)dW_j
\end{equation}
where \(x_i\) are some coordinates of pseudo-particles, \(t\) is the time, \(A_i\) is a drift and
\(B_{ij}\) a diffusion term. \(W_j\) is a Wiener process which can be related to a Gaussian 
distribution \(dw_j\) with mean zero and standard deviation of one \(N(0,1)\) by \(dW_j=\sqrt{dt}dw_j\).
The desired particle phase space distribution \(f\) is now obtained by solving the SDEs for a large sample of pseudo-particles. The SDEs for a simple one dimensional example are given by~\cite{yamada}
\begin{align}
\Delta r &= \left(- V + \frac{2\kappa_{rr}}{r}\right)\Delta t + \sqrt{2\kappa_{rr}\Delta t}dw_r,\\
\Delta T &= \frac{2V}{3r}\frac{T^2+2Tm}{T+m}\Delta t
\end{align}
with radial coordinate \(r\), particle mass \(m\) and kinetic energy \(T\). \(V\) labels a constant solar wind and \(\kappa_{rr}\) is the energy dependent diffusion constant. The initial condition at the boundary of the heliosphere which we typically set to \(r_{\text{Boundary}}=100 \text{ AU}\) for \(t=0\) is the local  interstellar flux. One can solve the SDEs in the so-called forward time approach. One starts with initial discrete kinetic energies \(T_i\) and simulates for small time steps \(\Delta t\) the evolution of \(T_i\) according to the SDE until either some large finite time or the position of the earth at \(r_{\text{Earth}}=1 \text{ AU}\) is reached. If the earth is reached, the final value \(T=T_i\) is recorded. Doing this for many pseudo-particles results in a continuous energy distribution for the same initial energy \(T_i\) of pseudo-particles. After normalization 
to one initial pseudo-particle this gives us a kind of Green's function \(G(T,T_i)\) which indicates for every initial energy \(T_i\) of some particle the probability to have kinetic energy \(T\) after reaching the earth (see e.g.~\cite{pei} for a more detailed discussion). 

\solarprop{} solves the SDEs in a different, so-called backward approach. Pseudo-particles start at the earth and are simulated backward in time until they reach the heliosphere. It has been shown that this approach is usually faster then the forward approach~\cite{kopp,DellaTorre:2015zgn}.
To incorporate the boundary condition, we weight the local interstellar flux \(\Phi^{\text{LIS}}(T)\) with the Green's function to obtain the flux at the top of the atmosphere \(\Phi^{\text{TOA}}(T)\). The flux is related to the phase space distribution by
\begin{equation}
\Phi^{\text{LIS}}(T)=p^2f^{\text{LIS}}(T)=(T^2+2Tm)f^{\text{LIS}}(T)
\end{equation}
and we get
\begin{equation}
f^{\text{TOA}}(T)=\sum_i G(T,T_i)f^{\text{LIS}}(T_i),\qquad \Phi^{\text{TOA}}(T)=p^2f^{\text{TOA}}(T).
\end{equation}
As the propagation depends on the mass of the cosmic rays, \solarprop{} calculates a Green's function for every given cosmic ray species independently. The main physical input is the definition of the solar magnetic field, the diffusion tensor and the treatment of the drift especially along the heliospheric current sheet (HCS). \solarprop{} has several different built-in models and can also easily extended by the user. 

\section{Installation}

\solarprop{} is freely available for download under~\cite{solarprop}. It needs the two packages \texttt{cfitsio} and \texttt{CCfits} to handle the FITS format. The packages can be downloaded and installed from 
\begin{code}
http://heasarc.gsfc.nasa.gov/fitsio/ 
\end{code}
and 
\begin{code}
http://heasarc.gsfc.nasa.gov/fitsio/ccfits/
\end{code}
Probably one has to tell \solarprop{} where these packages are installed on the system. 
After successful installation of the libraries, the following steps are necessary to install \solarprop{}. First, the code has to be extracted in a given favorite directory
\begin{code}
tar -xvf solarprop.tar.gz
\end{code}
Let us assume the packages \texttt{cfitsio} and \texttt{CCfits} are installed in \texttt{/usr/local/cfitsio} and \texttt{/usr/local/CCfits}.
There are two ways to tell \solarprop{} where the packages are. 
Either through environment variables or
during the \texttt{configure} step of the installation. 
\begin{enumerate}
\item
Using environment variables the steps are
\begin{code}
setenv LDFLAGS "-L/usr/local/cfitsio -L/usr/local/CCfits/lib"\\
setenv CPPFLAGS "-I/usr/local/cfitsio/include \verb$\$ \\
\hphantom{setenv CPPFLAGS "}-I/usr/local/CCfits/include"
\end{code}
for \texttt{csh} based systems or
\begin{code}
export LDFLAGS="-L/usr/local/cfitsio -L/usr/local/CCfits/lib"\\
export CPPFLAGS="-I/usr/local/cfitsio/include \verb$\$\\
\hphantom{export CPPFLAGS="}-I/usr/local/CCfits/include"
\end{code}
on \texttt{bash} like systems. Finally one has to run 
\begin{code}
./configure
\end{code}
\item
One can instead also use \texttt{configure} options and do
\begin{code}
./configure --with-CCfits=/usr/local/CCfits \verb$\$ \\
\hphantom{./configure }--with-cfitsio=/usr/local/cfitsio
\end{code}
\end{enumerate}
To install \solarprop{} in an individual directory one can use the \texttt{--prefix} option like 
\begin{code}
./configure --prefix=/local/solarprop --with-CCfits=/usr/local/CCfits \verb$\$ 
\\
\hphantom{./configure }--with-cfitsio=/usr/local/cfitsio
\end{code}
After that the next steps are 
\begin{code}
make\\
make install
\end{code}
to install \solarprop{}. The executable should be available in the subdirectory \texttt{./bin}. The simple call \texttt{./bin/SOLARPROP -v} should result in \texttt{This is SOLARPROP version 1.0} if the installation is successful.

\section{Usage of \solarprop{}}

After successful installation \solarprop{} can be used from the command line. There is no interactive user 
interface, everything is controlled with command line arguments and a control file. The program modulates a given or built in local 
interstellar flux and writes the computed top of the atmosphere (TOA) flux to a file. The physical model responsible for 
solar modulation is selected via a parameter in the control file. The local interstellar flux can be given either by a text
or FITS file or by the parametrization of the physical model. For an easy start we discuss different simple use cases.

\subsection{Use case 1: Input from \galprop{}}
\galprop{} stores the computed local interstellar flux of all defined cosmic ray species in a FITS file (see section \ref{sec:inout} or~\cite{galpropManual} for a description of the format). This file can directly be used as input for \solarprop{}. 
\begin{code}
./solarprop -c control.dat -o example -i nuclei\_54\_067f0001 -a angle.dat \verb$\$\\
\hphantom{./solarprop }-n nm.dat
\end{code}
The output of \solarprop{} is determined by the keywords in \texttt{control.dat}. An example for a control file is given below:
\begin{code}
model standard2D\\
outputFormat Both\\
modNumber 1
\end{code}
If this control file is used, \solarprop{} creates two output files, \texttt{example.dat} with text output and \texttt{example.fit} with the same data as FITS output. The keyword \texttt{modNumber} has the value 1, thus \solarprop{} calculates the modulation of the first cosmic ray species in the input FITS file from \galprop{}. All possible control file options are displayed in table \ref{tab:control1}, \ref{tab:control2}  and \ref{tab:options}.

\subsection{Use case 2: Input from \dragon{}} 
\dragon{} stores the computed local interstellar flux of the different cosmic ray species in a three dimensional FITS file, in a two dimensional FITS file and in a text file. \solarprop{} is able to handle the two dimensional FITS file. A brief description of the format can be found in section \ref{sec:inout}. A simple program call would be
\begin{code}
./solarprop -c control.dat -o example -i run\_2D\_spectrum.fits -n nm.dat
\end{code}
The output of \solarprop{} is determined by the keywords in \texttt{control.dat}. Consider the following control file:
\begin{code}
model standard2D\\
outputFormat Both\\
modNumber 1\\
polarity 1\\
angle 30
\end{code}
\solarprop{} creates two output files, \texttt{example.dat} with text output and \texttt{example.fit} with the same data as FITS output. The keyword \texttt{modNumber} has the value 1, thus \solarprop{} calculates the modulation of the first cosmic ray species in the FITS file from \dragon{}. The solar polarity and the tilt angle of the HCS are manually provided in the control file with the keywords \texttt{polarity} and \texttt{angle}. All possible control file options are displayed in table \ref{tab:control1}, \ref{tab:control2} and \ref{tab:options}.  

\subsection{Use case 3: Input from text file}

\solarprop{} can also handle input from text files. In this case the kinetic energy of the nuclei has to be given in GeV in the first column, the flux is expected in the second column. The mass and charge of the cosmic ray nuclei can be provided through the \texttt{mass} and \texttt{charge} keywords in the control file.
\begin{code}
./solarprop -c control.dat -o example -i textInput.dat -a angle.dat -n nm.dat
\end{code}
The output of \solarprop{} is determined by the keywords in \texttt{control.dat}. A possible control file is shown below:
\begin{code}
model standard2D\\
outputFormat FITS\\
mass 0.938\\
charge 1
\end{code}
Here \texttt{outputFormat} is given as FITS, thus \solarprop{} creates one output file \texttt{example.fit} with FITS output. The FITS file consists of two columns. The first one with the kinetic energy and the second one with the modulated flux. The keywords \texttt{mass} and \texttt{charge} specify the cosmic ray species to be protons.

\subsection{General use of \solarprop{}}
 
All possible command line arguments of \solarprop{} are given in table \ref{tab:command}.
\begin{table}[htb]
\centering
\begin{tabular}{lp{0.75\textwidth}}
Option&Description and argument\\
\hline
\texttt{-c, --control}&Argument is the file name where the control file is located.\\
\texttt{-a, --angle}&Argument is the file name where the tilt angle file is located. The format 
should match the definitions from~\cite{wsotilt}.\\
\texttt{-o, --output}&Argument is the file name where the output should be stored. The file ending
is added automatically depending on the option provided in the control file.\\
\texttt{-i, --input}&Argument is the file name where the input file is located. \solarprop{} automatically determines if the input is in FITS or Text format.\\
\texttt{-n, --neutron}&Argument is the file name where the file with the neutron monitor data is located. The file should be in the format from~\cite{phi}.\\
\texttt{-v, --version}&No argument necessary. If this option is chosen \solarprop{} only displays the version number. 
\end{tabular}
\caption{Command line options.}
\label{tab:command}
\end{table}
Everything else is controlled by the control file. Every line of the control file is interpreted as key value pair. The first word is a keyword, after a blank, an option should be provided. All keywords are listed in table \ref{tab:control1} and \ref{tab:control2}. There exists a default behavior of \solarprop{} for each keyword, for example in case no option is given, or the option is ill-defined. The possible options are given in table \ref{tab:options}.
\begin{table}[htb]
\centering
\begin{tabular}{lp{0.75\textwidth}}
Option&Description\\
\hline
\texttt{model}&Defines the physical model which should be used for solar modulation. Default is \texttt{ref1}.\\
\texttt{phi}&Force-field value in GV if provided. Result is stored in separate output. Default is no force-field approximation is computed.\\
\texttt{modNumber}&If provided number of cosmic ray species which should be used for solar modulation. If no value is provided, all cosmic ray species from the input file are modulated. Default is that all species are modulated.\\
\texttt{mass}&Mass of the cosmic ray species can be manually given. Is only considered if input is in Text format or no input is provided. Default is 
\(0.938\text{ GeV}\).\\
\texttt{charge}&Charge of the cosmic ray species can be manually given. Is considered if input is in Text format or no input is provided. Default
is 1.\\
\texttt{tiltModel}&Defines if R model or L model for the tilt angle should be used. Default is R model. Option is only used if the command line option \texttt{-a, --angle} is used.\\
\texttt{year}&Defines the start year for the determination of the tilt angle and the solar polarity. Average over the whole year is used if no month is given. Default is 2014.\\
\texttt{month}&Defines the start month for the determination of the tilt angle and the solar polarity. Average over the month is used if yearEnd and monthEnd are not provided. \\
\texttt{yearEnd}&Defines the end year for the determination of the tilt angle. Is considered if also monthEnd is provided.\\
\texttt{monthEnd}&Defines the end month for the determination of the tilt angle. Is considered if also yearEnd is provided. The average for the time period between month.year and monthEnd.yearEnd is determined. 
\end{tabular}
\caption{Control file options (Part 1).}
\label{tab:control1}
\end{table}
\begin{table}[htb]
\centering
\begin{tabular}{lp{0.75\textwidth}}
Option&Description\\
\hline
\texttt{polarity}&Defines solar polarity manually. Default is 1. Option is only used if the command line option \texttt{-a, --angle} is not used.\\
\texttt{angle}&Defines the tilt angle manually. Default is 0. Option is only used if the command line option \texttt{-a, --angle} is not used.\\
\texttt{outputFormat}&Defines the format for the output file. Default is \texttt{Text}.\\
\texttt{kappaScaling}&Normalization of the diffusion tensor \(\kappa\) can be manually adjusted. Default is 1.\\
\texttt{BfieldScaling}&Normalization of the magnetic field \(B\) can be manually adjusted. Default is 1.\\
\texttt{index}&Spectral index for output rescaling. Output flux is rescaled with \(T^{\texttt{index}}\). Default is no rescaling 0.\\
\texttt{dt}&Used time step \(\Delta t\) can be adjusted manually. Be careful with this option. A too large time step increases the speed, but results in wrong results. A too small time step results in a very high computing time.\\
\texttt{extraBins}&Number of extra bins which should be added between the given local interstellar spectrum data points. Computed bins are only added if the given data points from the input file are not dense enough. Default is 5.\\
\texttt{total}&Number of pseudo-particles which are modulated for every energy bin. Default is 1000.\\
\texttt{nmValue}&Value for the diffusion tensor normalization in the \texttt{standard2D} model. Default is 500.
Option is only used if the model \texttt{standard2D} is selected and command line option \texttt{-n, --neutron} is not used.
\end{tabular}
\caption{Control file options (Part 2).}
\label{tab:control2}
\end{table}

\begin{table}[htb]
\centering
\begin{tabular}{lp{0.75\textwidth}}
Keyword&Allowed parameters\\
\hline
\texttt{model}&\texttt{ref}, \texttt{ref2}, \texttt{ref3}, \texttt{ref4}, \texttt{standard2D} or \texttt{custom}.\\
\texttt{phi}&Every numerical value. Unit is GV.\\
\texttt{modNumber}&Integer value. Value should not be higher than the number of cosmic ray species in the input FITS file.\\
\texttt{mass}&Every numerical value. Unit is GeV.\\
\texttt{charge}&Integer value.\\
\texttt{tiltModel}&\texttt{L} or \texttt{R}.\\
\texttt{year}&Four digit integer value.\\
\texttt{month}&One or two digit integer value.\\
\texttt{yearEnd}&Four digit integer value.\\
\texttt{monthEnd}&One or two digit integer value.\\
\texttt{polarity}&\texttt{1} or \texttt{-1}.\\
\texttt{angle}&Every numerical value between 0 and 90. Unit is degree.\\
\texttt{outputFormat}&\texttt{FITS}, \texttt{Text} or \texttt{Both}.\\
\texttt{kappaScaling}&Every numerical value.\\
\texttt{BfieldScaling}&Every numerical value.\\
\texttt{index}&Every numerical value.\\
\texttt{dt}&Integer value.\\
\texttt{extraBins}&Integer value.\\
\texttt{total}&Integer value.\\
\texttt{nmValue}&Every numerical value.
\end{tabular}
\caption{Viable values for the keywords in the control file as described in table \ref{tab:control1} and \ref{tab:control2}.}
\label{tab:options}
\end{table}

A general program call is:
\begin{code}
./solarprop -c control.dat -a angle.dat -n nm.dat -o example -i input.fits
\end{code}
The two files \texttt{angle.dat} and \texttt{nm.dat} store the tilt angle and neutron monitor data which are necessary for the model \texttt{standard2D} (see the discussion in section \ref{sec:examples}). If another model is used, these files are not necessary. If one wants to adjust the result from \solarprop{}, e.g. to find a better agreement with a data set, the options \texttt{kappaScaling} and \texttt{BfieldScaling} can be used. The first one parametrizes the normalization of the diffusion, whereas the second one parametrizes the strength of the solar magnetic field. A larger value for \texttt{kappaScaling} e.g. \(1.5\) makes the diffusion more efficient and results in a higher top of the atmosphere flux.

If the option \texttt{phi} is given in the control file, \solarprop{} also calculates the result for the force-field approximation. This result is stored in a file with the name \texttt{exampleForce.dat} or \texttt{exampleForce.fit} depending on the value for \texttt{outputFormat}. 

\subsection{Input and output of \solarprop{}}
\label{sec:inout}

The local interstellar input flux for \solarprop{} can be provided through the command line option \texttt{-i} in three different ways.
\begin{enumerate}
\item If no input file is provided, \solarprop{} tries to find a reference local interstellar flux provided by the chosen model. If that is not successful, a warning is displayed and a vanishing local interstellar flux is assumed. In that case the output is useless. 
\item The local interstellar flux can be imported from a text file. The first column is assumed to contain the kinetic energy in GeV and the second column the local interstellar flux in \(\text{m}^{-2}\text{ sr}^{-1}\text{ s}^{-1}\text{ GeV}^{-1}\). The mass and charge of the cosmic ray species have to be given in the control file.
\item If a FITS file is given, \solarprop{} checks automatically if the file is in \texttt{GALPROP} or \texttt{DRAGON} like format. Other formats are not supported. In case a \texttt{modNumber} is given, only the desired cosmic ray species is modulated. If no \texttt{modNumber} is found, \solarprop{} modulates all cosmic ray species found in the import file.
\end{enumerate}

The output of \solarprop{} is stored in files which names are provided through the option \texttt{-o}. The output depends on the keywords \texttt{modNumber} and \texttt{outputFormat} and on the format of the input. 
We can distinguish between two cases.
\begin{enumerate}
\item If the input is from a text file or the local interstellar flux is from an internal reference model, independent of the \texttt{modNumber} only one cosmic ray species is modulated as only one is provided. In the case of text file output, in the first column the kinetic energy in GeV is written. In the second column the top of the atmosphere flux is provided in units of \(\text{m}^{-2}\text{ sr}^{-1}\text{ s}^{-1}\text{ GeV}^{-1}\) (if \texttt{index} is 0). The energy bins are always the same as the input ones from the local interstellar flux.

In case of FITS output, two keywords, namely \texttt{mass} and \texttt{charge} to store the cosmic ray properties are added to the pHDU. The first column of the pHDU consists of the kinetic energy in GeV and the second column stores the calculated flux in \(\text{m}^{-2}\text{ sr}^{-1}\text{ s}^{-1}\text{ GeV}^{-1}\) (if \texttt{index} is 0).
\item If the input is from a FITS file, depending on the value of the keyword \texttt{modNumber} one or all cosmic ray species in the file are modulated. All species are present in the output. If the species is not modulated the local interstellar flux is added to the output. In the case of text output, the file consists of one column with the kinetic energy in GeV and additional columns, one for the flux of each cosmic ray species in \(\text{m}^{-2}\text{ sr}^{-1}\text{ s}^{-1}\text{ GeV}^{-1}\) (if \texttt{index} is 0).

For FITS output, \solarprop{} uses the same format as given by the input. That means, if the input is in \texttt{GALPROP} like format the fluxes are stored, each in a separate column in the pHDU, whereas for a \texttt{DRAGON} like format every flux is stored in its own HDU. Please note that the unit of the stored fluxes in both cases is changed to the \solarprop{} standard \(\text{m}^{-2}\text{ sr}^{-1}\text{ s}^{-1}\text{ GeV}^{-1}\)!
\end{enumerate} 

\subsection{Inclusion of custom models}

\solarprop{} can be easily extended by the user. The main classes for the physics of particle propagation are the interface (abstract base class) \texttt{Iparticle.h}, the derived abstract class \texttt{particle.cc} and its concrete implementations like \texttt{standard2D.cc}. If one wants to include an own model, it should inherit from \texttt{particle.cc}. Of course, all virtual methods of \texttt{particle.cc} have to be implemented and a constructor has to be added. The method \texttt{void calculate()} manages the computation of all quantities which are needed several times in a propagation step. This avoids unnecessary double computations. A basic implementation can already be found in \texttt{particle.cc}.

\solarprop{} already offers two files for the concrete implementation of a custom model, \texttt{custom.h} and \texttt{custom.cc}. The implemented model can be called from the control file with the name \texttt{custom} as described for the validation in \texttt{import.h}. 

\section{Examples and validation}

\subsection{Validation}

\begin{figure}[htb]
\centering
\includegraphics[width=7.5cm]{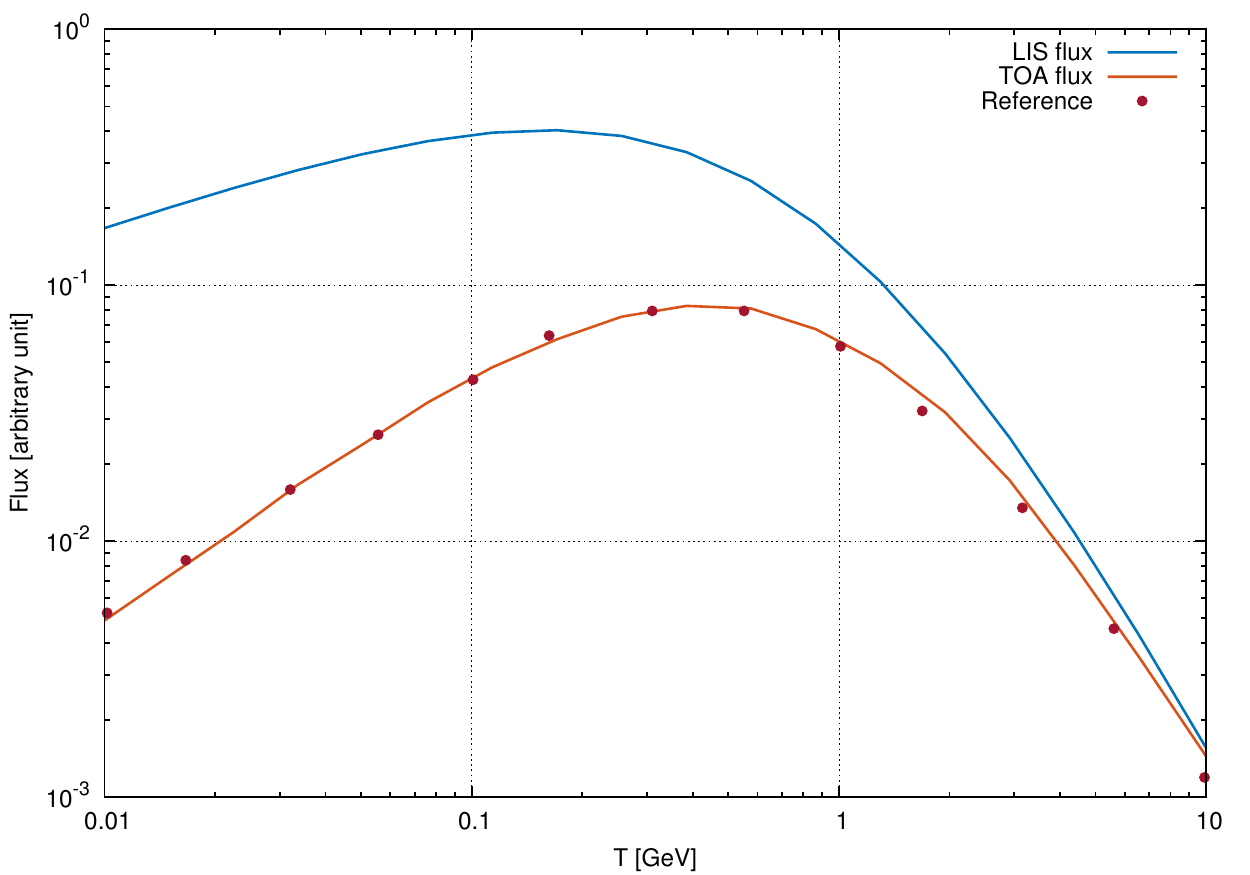}
\includegraphics[width=7.5cm]{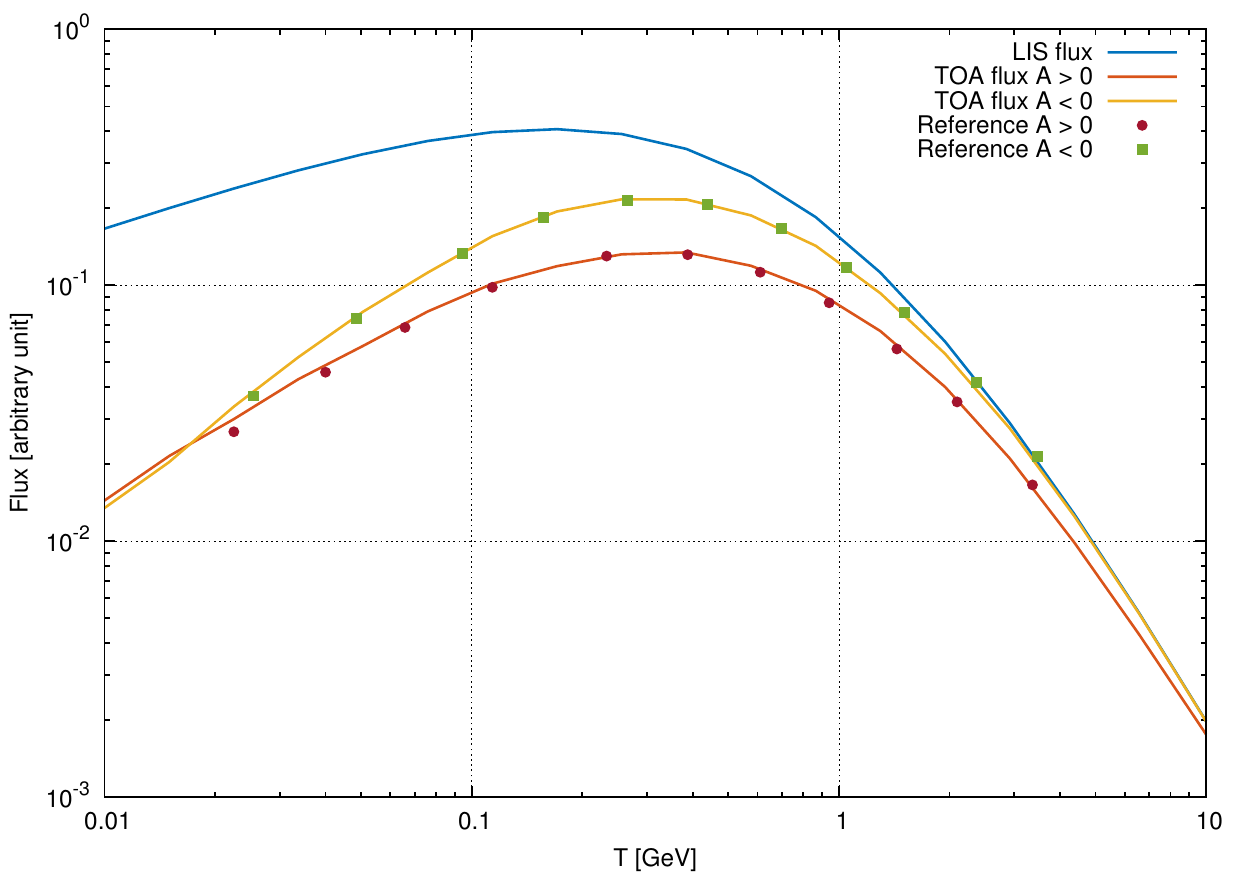}
\caption{Validation of \solarprop{} with the reference model \texttt{ref1} from~\cite{yamada} (left panel) and the model \texttt{ref2} from~\cite{jokipii} (right panel).}
\label{fig:yamada}
\end{figure}

\begin{figure}[htb]
\centering
\includegraphics[width=7.5cm]{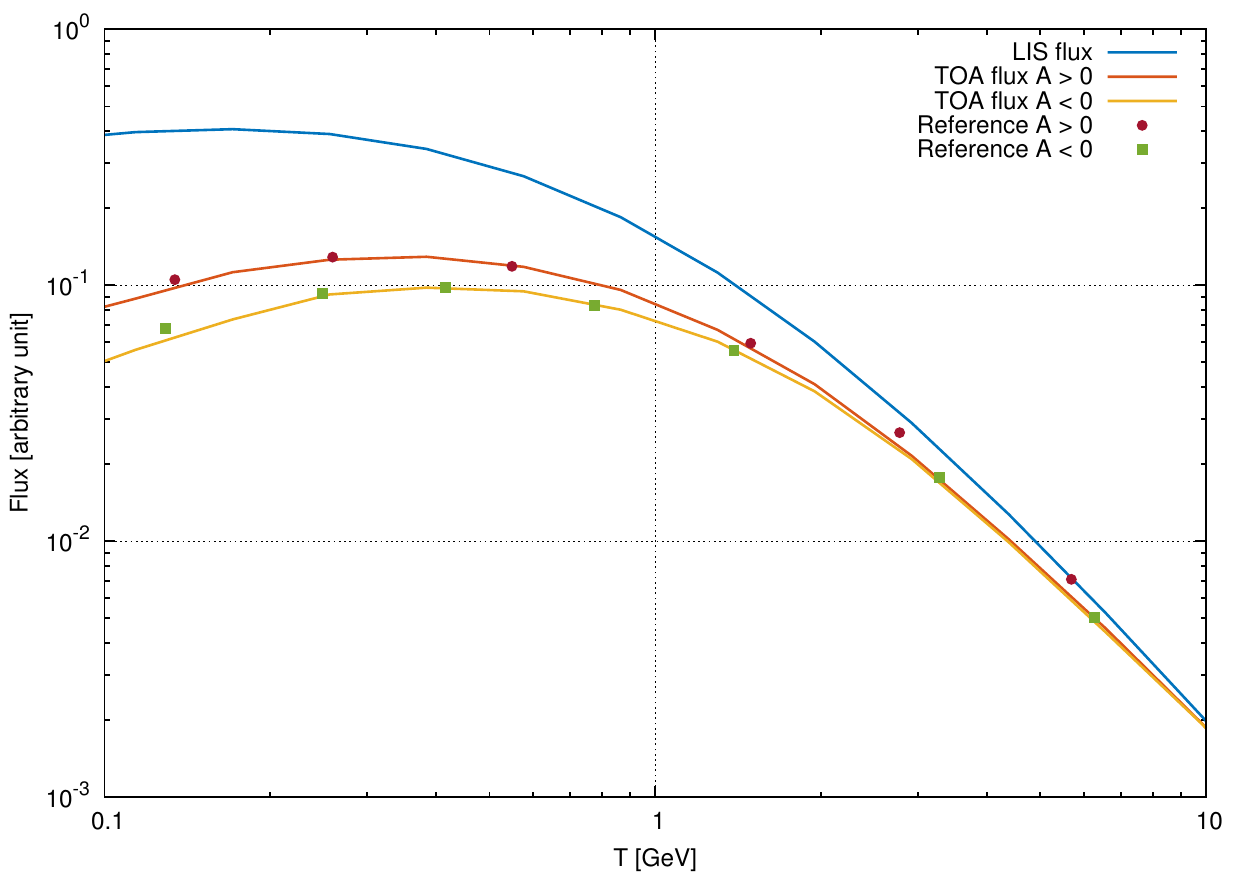}
\includegraphics[width=7.5cm]{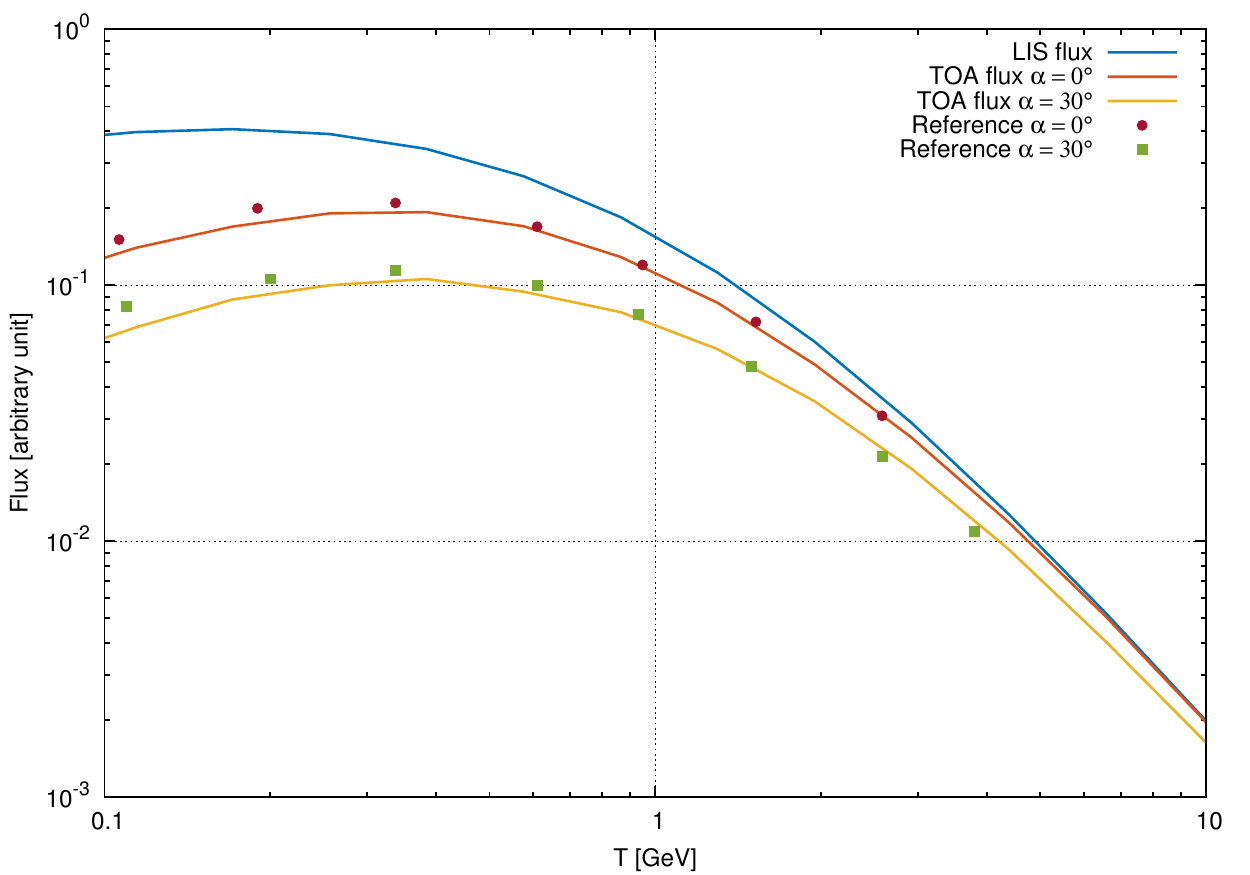}
\caption{Validation of \solarprop{} with the reference model \texttt{ref3} from~\cite{potgieter} (left panel) and the model \texttt{ref4} from~\cite{burger} (right panel).}
\label{fig:burger}
\end{figure}

We validate \solarprop{} against several models available in the literature. This approach has also been used previously to test Monte Carlo based solar modulation calculations~\cite{pei}. A comparison of the result from \solarprop{} with the result of~\cite{yamada} and~\cite{jokipii} is shown in figure \ref{fig:yamada}. In the left panel a one dimensional model with plain diffusion and a diffusion coefficient \(\kappa =5\cdot 10^{22} \text{ cm}^2 \text{ s}^{-1} \text{ GV}^{-1}\) is used (matches figure 4 in~\cite{yamada}). The model is implemented for reference in \solarprop{} as model \texttt{ref1}. The corresponding SDEs can be found in appendix \ref{sec:ref1}. The control file to reproduce this result is very simple:
\begin{code}
model ref1\\
mass 0.938\\
charge 1
\end{code}
The corresponding program call is:
\begin{code}
./solarprop -c control.dat -o example
\end{code}
In the right panel a two dimensional model without tilt angle dependence (flat HCS) is used. The parameters match the ones in table 1 and figure 2 of~\cite{jokipii}. The model can be used in \solarprop{} with parameter \texttt{ref2} and the SDEs can be found in appendix \ref{sec:ref2}. For negative polarity \(A<0\) the control file to reproduce the figure is:
\begin{code}
model ref2\\
mass 0.938\\
charge 1\\
polarity -1
\end{code}
A comparison of more sophisticated models is displayed in figure \ref{fig:burger}. In the left panel a two dimensional model with a possibility to model a wavy HCS is used~\cite{potgieter}. The plot displays a comparison between \solarprop{} and the result from figure 6 of~\cite{potgieter}. The SDEs can be found in appendix \ref{sec:ref3}. The corresponding control file for positive polarity \(A>0\) is:
\begin{code}
model ref3\\
mass 0.938\\
charge 1\\
polarity 1
\end{code} 
In the right panel the result of figure 5 from~\cite{burger} is reproduced for two different values of the tilt angle \(\alpha\). The model is also two dimensional and able to describe a wavy HCS. As can be seen in appendix \ref{sec:ref4} the drifts are described in a different way than in~\cite{potgieter}. The following options have been used to produce the result for the case \(A<0\) and \(\alpha = 30^\circ\).
\begin{code}
model ref4\\
mass 0.938\\
charge 1\\
polarity -1\\
angle 30
\end{code}
The models of figure \ref{fig:burger} are available in \solarprop{} under the name \texttt{ref3} and \texttt{ref4} respectively.
The results from figure \ref{fig:yamada} and figure \ref{fig:burger} nicely show that our Monte Carlo approach is viable and different models of the heliosphere can be easily implemented in \solarprop{}.

\subsection{Examples}
\label{sec:examples}

\begin{figure}[htb]
\centering
\includegraphics[width=12.0cm]{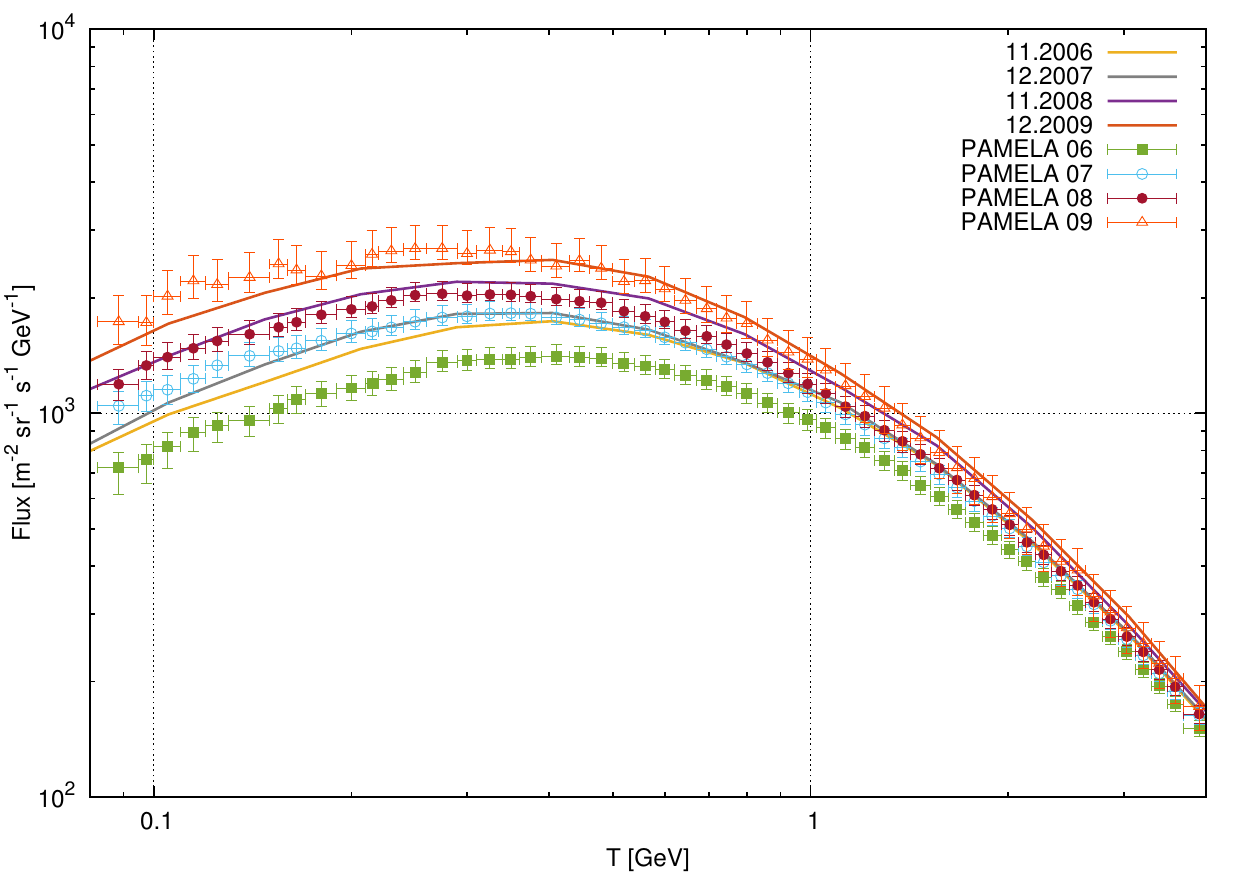}
\caption{Data for cosmic ray protons at different solar activities measured by PAMELA~\cite{Adriani:2013as}. The result from \solarprop{} with the model \texttt{standard2D} is displayed as solid lines. The result is in good agreement, except for the data from 2006.}
\label{fig:pamela}
\end{figure}

We want to introduce in this section another model which is implemented in \solarprop{} and show its result with different data sets.
This more sophisticated model, able to describe recent cosmic ray data is implemented in \solarprop{} under the name \texttt{standard2D}. The model is still very simple, but has several time dependent parameters to accurately describe recent data. A more complicated model, e.g. a three dimensional one, can easily added by the user or will maybe be part of the next version of \solarprop{}. A challenging data set for cosmic ray protons released by the PAMELA experiment~\cite{Adriani:2013as} is displayed in figure \ref{fig:pamela}.
The proton flux increases with time which indicates a decreasing solar activity. This data set has been successfully described by solar modulation models~\cite{Potgieter:2013cwj,Raath:2015zga}. These studies have shown that a simple model, where only one parameter (e.g. the tilt angle) is time dependent is insufficient to describe the data. This excludes all our discussed reference models. In table \ref{tab:pamelaData} several measured solar activity dependent parameters are displayed. 
\begin{table}[hbt!]
\centering
\begin{tabular}{ccccc}
Time&\(\alpha_L\)&\(\alpha_R\)&SSN&NM count rate (Newark)\\
\hline
11.2006&29.5&11.25&12.6&3539\\
12.2007&31.267&15.8&4.9&3591\\
11.2008&26&9.85&1.7&3678\\
12.2009&28.3&13.55&8.3&3732
\end{tabular}
\caption{Different solar activity dependent parameters as measured by~\cite{hoeksema,wsotilt,ssn,nm} for a time of negative solar polarity \(A < 0\). The increasing NM count rate indicates a decreasing solar activity.}   
\label{tab:pamelaData}
\end{table}
To accurately describe experimental data a tilt angle and sunspot number (SSN) dependent model was used in~\cite{Bobik:2011ig}. In~\cite{Potgieter:2013cwj,Raath:2015zga} many more parameters, like the overall magnetic field strength  have been adjusted. We use a tilt angle dependent model and vary in addition the normalization of the diffusion tensor \(\kappa_0\), taking into account a more or less effective diffusion. 

The variation of \(\kappa_0\) is based on the calculations of~\cite{usoskin05,usoskin11} which average different neutron monitor (NM) count rate data. In~\cite{usoskin05,usoskin11} a force-field potential \(\phi_{\text{Usoskin}}\) was determined from the neutron monitor data to describe charge-sign independent solar modulation. We relate the diffusion tensor normalization \(\kappa_0\) to this quantity \(\phi_{\text{Usoskin}}\). By not directly using the neutron monitor data we avoid to redo the average procedure, already performed by~\cite{usoskin05,usoskin11}. We use the simple relation 
\begin{equation}
\label{equa:kappatime}
\kappa_0=\tilde{\kappa_0}
\begin{cases}
\frac{137}{\phi_{\text{Usoskin}}}- 0.061&, qA <0\\
\frac{7}{100}\frac{137}{\phi_{\text{Usoskin}}}- 0.061&, qA >0
\end{cases}
\end{equation}   
to model a solar activity dependent diffusion tensor normalization. \(\tilde{\kappa_0}\) is just the usual 
diffusion constant.
The formula depends on the polarity of the solar cycle \(A\) and the charge of the cosmic ray species \(q\) to take into account that drift effects have not been considered in~\cite{usoskin05,usoskin11}. This idea is also motivated by the approach in~\cite{Bobik:2011ig} where the normalization of the diffusion tensor is polarity and charge dependent. The detailed implementation of the drift effects is described in appendix \ref{sec:standard}. To reproduce the results in figure \ref{fig:pamela} e.g. for the date 11.2007 the control file is:
\begin{code}
model standard2D\\
month 11\\
year 2007\\
mass 0.938\\
charge 1
\end{code}
Except for the data from 2006 the model \texttt{standard2D} is in good agreement with the data. To describe all four data sets a more complicated model is necessary. 

\begin{figure}[htb]
\centering
\includegraphics[width=12.0cm]{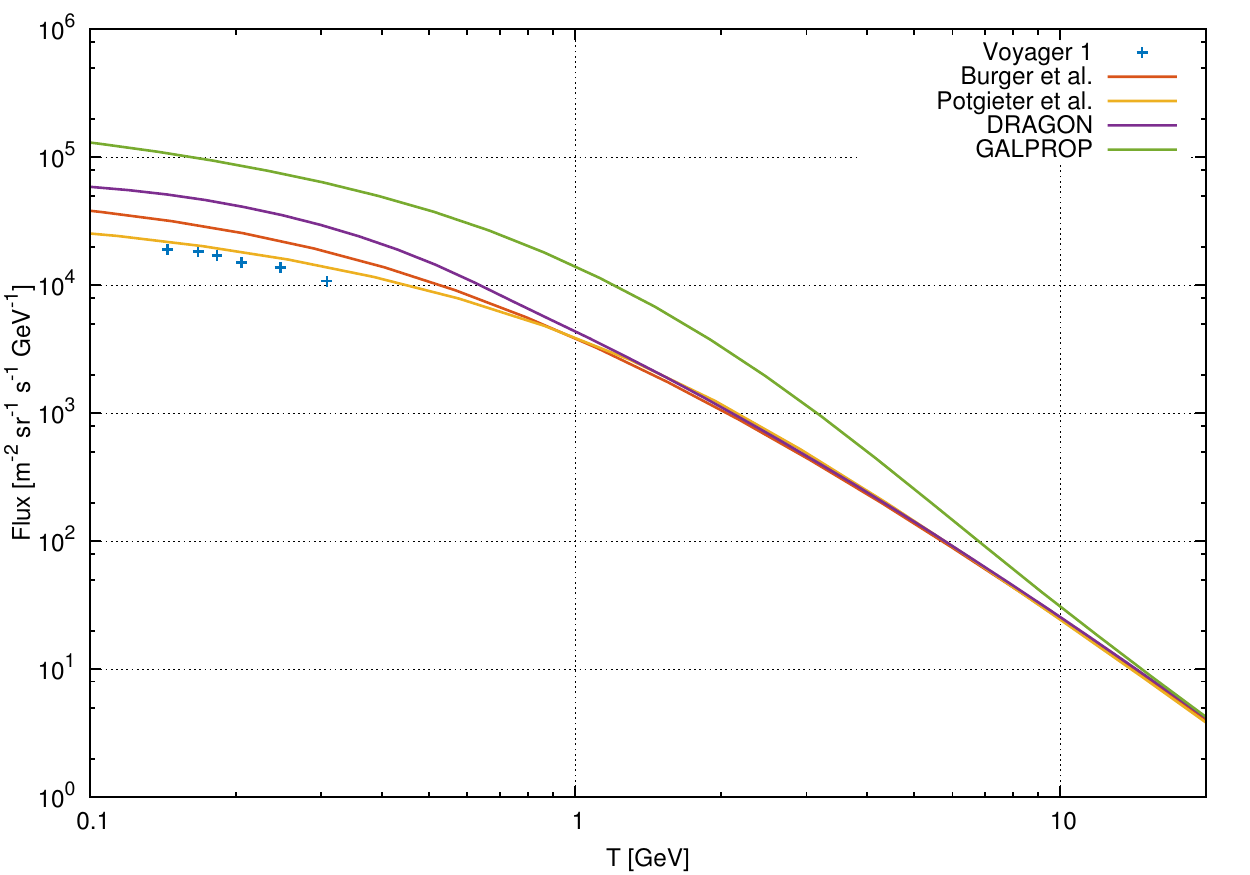}
\caption{Voyager 1 measurements for protons at 122 AU~\cite{voyager} and different popular local interstellar proton fluxes~\cite{burger2000,Potgieter:2013cwj} including results from \galprop{} and \dragon{}. The fluxes from \galprop{} and \dragon{} are of course model dependent.}
\label{fig:lis}
\end{figure}

We want to point out that the concrete form of a successful model for solar modulation strongly depends on the local interstellar flux. Obviously, a different shape of the diffusion tensor or a different approach to model the drift effects is possible if the local interstellar flux is adjusted. As the Voyager 1 spacecraft has recently passed the heliosphere and entered the heliopause~\cite{voyager}, the local interstellar flux for some cosmic ray species is now measured directly for the first time. We show this data in figure \ref{fig:lis} together with different local interstellar proton fluxes. 
We take the local interstellar proton flux from~\cite{burger2000} as also used in~\cite{usoskin05,usoskin11,Bobik:2011ig} (notice the comment in~\cite{usoskin05} about the different forms of the equation). As figure \ref{fig:lis} indicates, the flux from~\cite{Potgieter:2013cwj} seems to be more appropriate. Because we use results from~\cite{usoskin05, usoskin11}, we restrict ourselves to the proton flux from~\cite{burger2000}. This local interstellar flux is used for the result displayed in figure \ref{fig:pamela}.

\begin{figure}[htb]
\centering
\includegraphics[width=7.5cm]{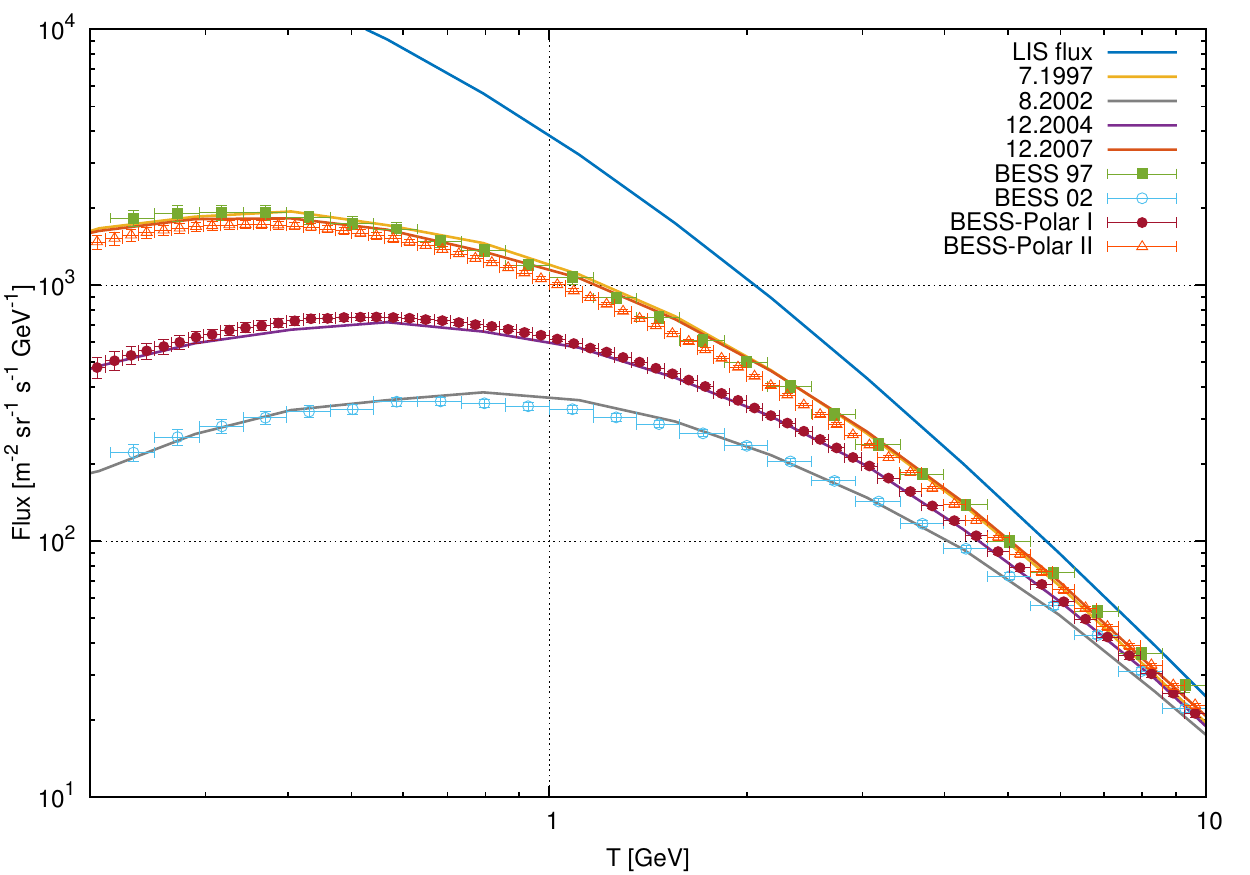}
\includegraphics[width=7.5cm]{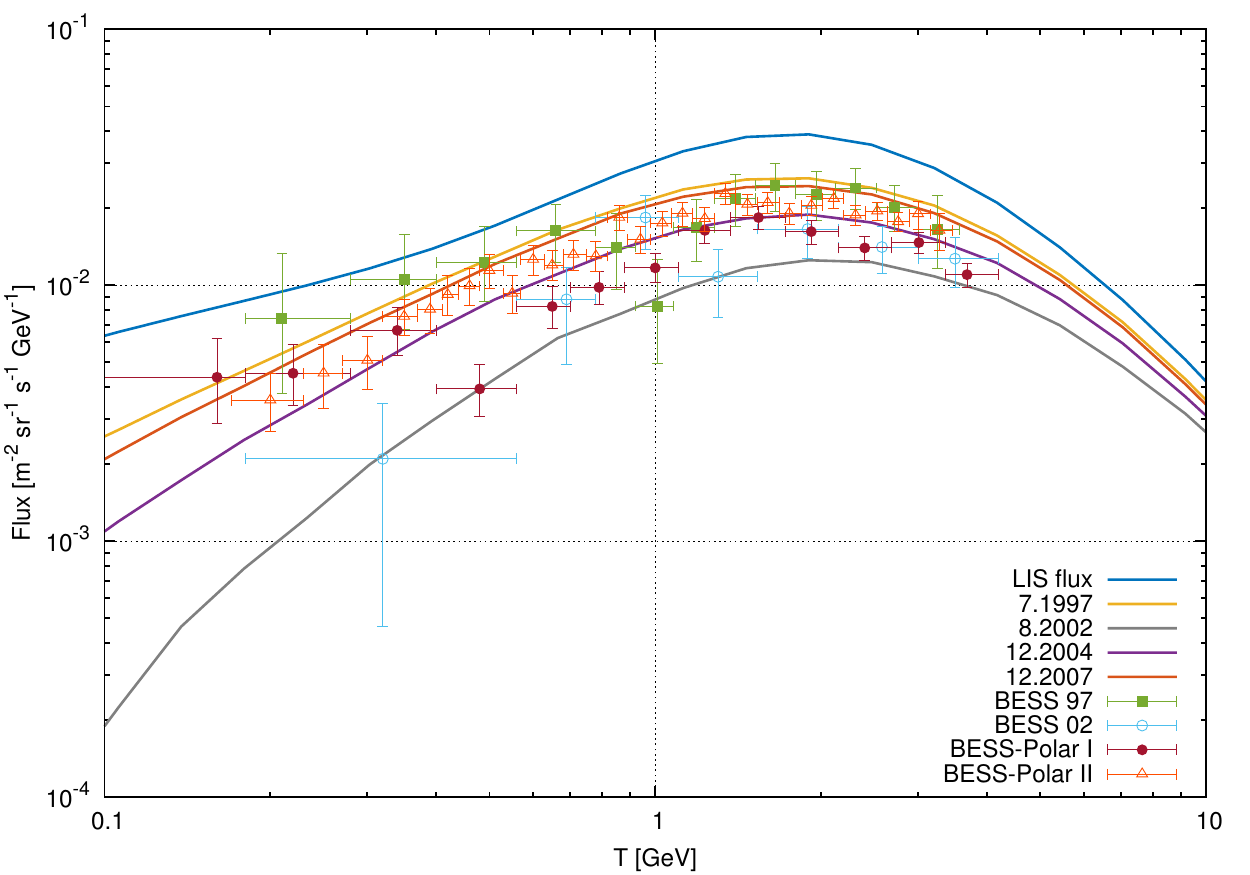}
\caption{Data for cosmic ray protons and antiprotons at different solar activities measured by BESS 97, BESS 2002~\cite{Orito:1999re,bess2002,Shikaze:2006je} and BESS-Polar I, BESS-Polar II~\cite{Abe:2008sh,Abe:2011nx,Abe:2015mga}.}
\label{fig:besspolar}
\end{figure}

In figure \ref{fig:besspolar} the result for cosmic ray modulation with the \texttt{standard2D} model of \solarprop{} and data from the BESS and BESS-Polar experiments~\cite{Orito:1999re,bess2002,Shikaze:2006je,Abe:2008sh,Abe:2011nx,Abe:2015mga} are displayed. The local interstellar flux for antiprotons is determined with recent results from the literature.
The galactic propagation setup described in~\cite{Kappl:2015bqa} together with the recently updated antiproton production cross sections~\cite{Kappl:2014hha} and the proton flux from~\cite{burger2000} is used. Emphasizing that the model uses no fit parameters it describes the data very well. The only input is the time when the experiment took place.
The result slightly overshoots the data from the BESS-Polar II experiment which took data in December 2007. As there is a deviation between the PAMELA proton data from the same time period and the BESS-Polar II data we do not worry about this result.

\begin{figure}[htb]
\centering
\includegraphics[width=12.0cm]{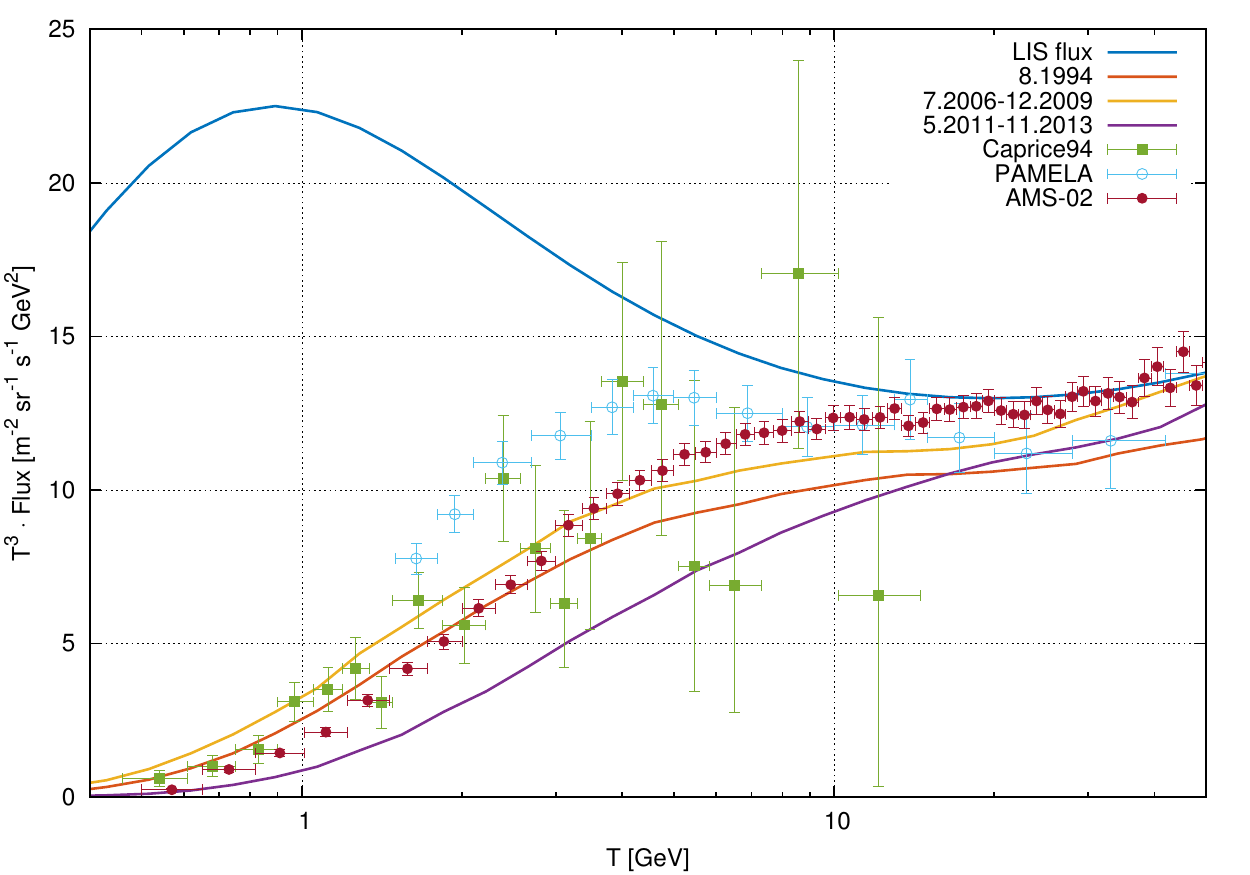}
\caption{Data for cosmic ray positrons at different solar activities measured by Caprice~\cite{caprice}, PAMELA~\cite{Adriani:2013uda} and AMS-02~\cite{Aguilar:2014mma}. The model \texttt{standard2D} is in poor agreement with the data. The question if this indicates that the solar modulation model is too simple or the local interstellar positron flux is different is left for future work.}
\label{fig:leptons}
\end{figure}

One can also describe leptons as can seen in figure \ref{fig:leptons}. 
This data (especially the positron fraction) has also been discussed in~\cite{Maccione:2012cu} as indication for charge-sign dependent solar modulation. The local interstellar positron flux seems to include an exotic primary contribution at higher energies, known as the positron excess (see e.g.~\cite{DiMauro:2015jxa} for a recent discussion of the possible primary positron contribution). 

For the parametrization we use the secondary positron flux from \dragon{} \(\Phi^{\text{LIS}}_{\text{sec, }\texttt{DRAGON}}(T)\) and add a power law contribution \(\Phi^{\text{LIS}}_{\text{prim}}(T)=3.7\cdot T^{-2.7}\) to model the primary contribution. This flux choice 
\(\Phi^{\text{LIS}}_{e^+}(T)=\Phi^{\text{LIS}}_{\text{sec, }\texttt{DRAGON}}(T)+\Phi^{\text{LIS}}_{\text{prim}}(T)\)
is in agreement with the high energy data by PAMELA~\cite{Adriani:2013uda} and AMS-02~\cite{Aguilar:2014mma}.
The result for solar modulation with this local interstellar flux and the model \texttt{standard2D} is in poor agreement with the data. This shows that we need a refined model to describe also solar modulation for leptons in a reasonable way or the local interstellar flux for positrons is different. It has recently been shown that more sophisticated models~\cite{Potgieter:2015jxa} can well describe new electron data by the PAMELA experiment~\cite{Adriani:2015kxa}. The inclusion of such models which seem to be able to describe solar modulation for leptons in \solarprop{} is left for future work. To reproduce e.g. the theoretical estimate for the positron flux of the AMS-02 experiment displayed in figure \ref{fig:leptons} the control file is:
\begin{code}
model standard2D\\
year 2011\\
month 5\\
yearEnd 2013\\
monthEnd 11\\
mass 0.000511\\
charge 1\\
index 3
\end{code}
together with the input flux as described above.

The examples in this section show that the simple \texttt{standard2D} model implemented in \solarprop{} is able to describe charge-sign dependent solar modulation. More sophisticated models which better describe the data can be implemented by the user.

\section{Conclusions}

We have introduced a tool called \solarprop{} to simulate charge-sign dependent solar modulation for cosmic rays. This is an advantage to the commonly used force-field approximation, as also drift effects which are charge-sign dependent are taken into account. We have validated the tool against several simple models from the literature, to show that the software works as expected. A more sophisticated two dimensional model has also been implemented. The only required input for this model is a date or a time period. With this information and a local interstellar flux, \solarprop{} calculates the top of the atmosphere flux. The results have been compared with several proton and antiproton data sets from BESS, BESS-Polar and PAMELA and are in a good agreement. The comparison with positron data from PAMELA and AMS-02 is more challenging. We find some tension with the data, which probably show that the implemented model is too simple to account for solar modulation for leptons. On the other hand also the local interstellar flux for positrons is rather uncertain due to its possible primary component. 

\solarprop{} has the advantage that it can be easily extended by new models. The implementation of custom models will make it possible to test further models for leptons. This is left for future work. Another advantage of \solarprop{} is its FITS interface. With this interface, output from popular tools for the propagation of cosmic rays in the galaxy like \galprop{} and \dragon{} can directly used as input for \solarprop{}. In the past, the free Fisk potential in the force-field method was often degenerate with fit parameters from galaxy propagation of cosmic rays like the strength of reacceleration. We hope that \solarprop{} can help to break this degeneracy and will lead in the combination with a detailed study of cosmic ray transport in the galaxy to a smaller propagation uncertainty at GeV kinetic energies. This is important for indirect dark matter detection with antiprotons and positrons and will hopefully be useful in the near future.

\subsection*{Acknowledgments}

This work was supported by the SFB-Transregio TR33 "The Dark Universe" 
(Deutsche Forschungsgemeinschaft)
and by the Munich Institute for Astro- and
Particle Physics (MIAPP) of the DFG cluster of excellence "Origin and
Structure of the Universe".

\appendix

\section{Description of the models}

We summarize the different SDEs which are implemented for the different models in \solarprop{}. The rigidity for a particle with momentum \(p\), mass \(m\) and charge \(Z\) is defined as
\begin{equation}
\mathcal{R}=\frac{p}{|Z|}=\frac{\sqrt{T^2+2Tm}}{|Z|}.
\end{equation}

\subsection{One dimensional models}

\subsubsection{Model \texttt{ref1}}
\label{sec:ref1}

The one dimensional model \texttt{ref1}~\cite{yamada} is described by
\begin{align}
\Delta r &= \left(- V + \frac{2\kappa_{rr}}{r}\right)\Delta t + \sqrt{2\kappa_{rr}\Delta t}dw_r,\\
\Delta T &= \frac{2V}{3r}\frac{T^2+2Tm}{T+m}\Delta t.
\end{align}
The solar wind is assumed to be constant \(V=400\text{ km}\text{ s}^{-1}\) and \(dw_j\) is a 
Gaussian 
distribution of random numbers with mean zero and standard deviation of one \(N(0,1)\).
The radial part of the diffusion tensor is given by
\begin{equation}
\kappa_{rr}=\kappa_0\beta\mathcal{R}=\kappa_0\beta\frac{\sqrt{T^2+2Tm}}{|Z|}.
\end{equation}
\(\beta\) is the particle speed which can be linked to the kinetic energy.

\subsection{Two dimensional models}

We assume for all models that the heliospheric magnetic field is described by a Parker spiral (see~\cite{Raath:2015zga} for a recent discussion of different magnetic fields)
\begin{equation}
\mathbf{B} =\frac{A}{r^2}(\mathbf{e_r}-\tan\psi \mathbf{e_{\phi}})
\left(1-2H\left(\vartheta-\frac{\pi}{2}\right)\right)
,\qquad B=\frac{|A|}{r^2}\sqrt{1+\tan^2\psi}.
\end{equation}
\(A\) is a constant taking care that the magnetic field at the position of the earth matches its measured value \(\approx 5\text{ nT}\) and the correct polarity of the solar cycle. We thus have usually \(A\approx\pm 3.4\text{ nT}\text{ AU}^{2}\). The factor \(\pi / 2\) in the Heaviside function is strictly valid only for a flat HCS and has to be corrected otherwise.
The spiral angle \(\psi\) is defined as
\begin{equation}
\tan\psi=\Gamma=\frac{\Omega r \sin\vartheta}{V}
\end{equation}
where \(\Omega = 2.866\cdot 10^{-6} \text{ rad}\text{ s}^{-1}\) is the average angular rotation of the sun. The radial and angular part of the diffusion tensor can be rewritten~\cite{burger2008}
\begin{align}
\kappa_{rr}&=\kappa_{\parallel}\cos^2\psi+\kappa_{\bot}\sin^2\psi
=\frac{1}{1+\Gamma^2}(\kappa_{\parallel}+ \kappa_{\bot}\Gamma^2),\\
\kappa_{\vartheta\vartheta}&=\kappa_{\bot}.
\end{align}
For the SDEs we find
\begin{align}
\Delta r &= \left(- V - V_{\text{D},r} - V_{\text{HCS},r} + \frac{1}{r^2}\frac{\partial r^2\kappa_{rr}}{\partial r}
\right)\Delta t + \sqrt{2\kappa_{rr}\Delta t}dw_r,\\
\Delta \vartheta &=\left(-\frac{V_{\text{D},\vartheta}}{r}+\frac{1}{r^2\sin\vartheta}\frac{\partial
\sin\vartheta\kappa_{\vartheta\vartheta}}{\partial \vartheta} \right)
\Delta t+ \frac{\sqrt{2\kappa_{\vartheta\vartheta}\Delta t}}{r}dw_{\vartheta},\\
\Delta T &= \frac{2V}{3r}\frac{T^2+2Tm}{T+m}\Delta t.
\end{align}

\subsubsection{Model \texttt{ref2}}
\label{sec:ref2}

The two dimensional model \texttt{ref2}~\cite{jokipii} is described by a \(r\) independent diffusion tensor 
\begin{align}
\kappa_{\parallel}&=\kappa_0\beta\sqrt{\mathcal{R}}=\kappa_0\beta\frac{(T^2+2Tm)^{\frac{1}{4}}}{\sqrt{|Z|}},\\
\kappa_{\bot}&=0.1\kappa_{\parallel}
\end{align}
which results in
\begin{align}
\frac{1}{r^2}\frac{\partial r^2\kappa_{rr}}{\partial r}&=
\frac{2}{r}\left(\kappa_{rr}+(\kappa_{\bot}-\kappa_{\parallel})\frac{\Gamma^2}{(1+\Gamma^2)^2}
\right),
\\
\frac{1}{r^2\sin\vartheta}\frac{\partial
\sin\vartheta\kappa_{\vartheta\vartheta}}{\partial \vartheta}&=
\frac{\kappa_{\bot}}{r^2}\cot\vartheta.
\end{align}
The HCS is assumed to be flat. To avoid a singular drift velocity due to the flat HCS we use a regularization proposed in~\cite{burger85}. In~\cite{jokipii} a different approach is used, but our validation shows that the result does not depend on the details of the regularization. The drift velocities are given by
\begin{align}
V_{\text{D},r}&=q\frac{2\beta r}{3A}\sqrt{T^2+2Tm}\cot\vartheta
\frac{\Gamma}{(1+\Gamma^2)^2}\left(1-2H\left(\vartheta-\frac{\pi}{2}\right)\right),\\
V_{\text{D},\vartheta}&=
q\frac{2\beta r}{3A}\sqrt{T^2+2Tm}
\frac{\Gamma(2+\Gamma^2)}{(1+\Gamma^2)^2}\left(1-2H\left(\vartheta-\frac{\pi}{2}\right)\right),\\
V_{\text{HCS},r}&=
\begin{cases}
-q\frac{A\beta}{|A|}\frac{\Gamma}{\sqrt{1+\Gamma^2}}\left(0.457-0.412\frac{d}{R_L}+0.0915\frac{d^2}{R_L^2}\right)&, d < 2 R_L\\
0&, d\geq 2 R_L.
\end{cases}
\end{align}
where the Larmor radius \(R_L\) and the distance to the HCS \(d\) are given by 
\begin{align}
R_L&=\frac{\mathcal{R}}{B}=\frac{\sqrt{T^2+2Tm}}{|Z|}\frac{r^2}{|A|\sqrt{1+\Gamma^2}},\\
d&=|r\cos\vartheta|.
\end{align}

\subsubsection{Model \texttt{ref3}}
\label{sec:ref3}

Another two dimensional model is the one named \texttt{ref3}~\cite{potgieter}. We find for the diffusion tensor
\begin{align}
\kappa_{\parallel}&=
\begin{cases}
\kappa_{0\parallel}\beta\sqrt{0.4}(1+r^2)&, \mathcal{R} < 0.4 \text{ GV}\\
\kappa_{0\parallel}\beta\sqrt{\mathcal{R}}(1+r^2)=\kappa_{0\parallel}\beta\frac{(T^2+2Tm)^{\frac{1}{4}}}{\sqrt{|Z|}}(1+r^2)&, \mathcal{R} \geq 0.4 \text{ GV},
\end{cases}\\
\kappa_{\bot}&=\kappa_{0\bot}\beta\mathcal{R}\frac{r^2}{\sqrt{1+\Gamma^2}}
=\kappa_{0\bot}\beta\frac{\sqrt{T^2+2Tm}}{|Z|}\frac{r^2}{\sqrt{1+\Gamma^2}}
\end{align}
which gives us
\begin{align}
\frac{1}{r^2}\frac{\partial r^2\kappa_{rr}}{\partial r}&=
\frac{2}{r}\left(\kappa_{rr}+(\kappa_{\bot}-\kappa_{\parallel})\frac{\Gamma^2}{(1+\Gamma^2)^2}
+\frac{r^2}{1+r^2}\frac{\kappa_{\parallel}}{1+\Gamma^2}
+\kappa_{\bot}\frac{\Gamma^2(2+\Gamma^2)}{2(1+\Gamma^2)^2}
\right),
\\
\frac{1}{r^2\sin\vartheta}\frac{\partial
\sin\vartheta\kappa_{\vartheta\vartheta}}{\partial \vartheta}&=
\frac{1}{1+\Gamma^2}\frac{\kappa_{\bot}}{r^2}\cot\vartheta.
\end{align}
The model simulates a wavy HCS and introduces a smooth function \(f\) and its derivative \(f'\) to model the HCS drifts.
We find
\begin{align}
f&=\frac{1}{\alpha_{\text{HCS}}}\arctan\left(\left(1-\frac{2\vartheta}{\pi}\right)\tan\alpha_{\text{HCS}}\right),\\
f'&=-\frac{2}{\pi\alpha_{\text{HCS}}}\frac{\tan\alpha_{\text{HCS}}}{1+\left(1-\frac{2\vartheta}{\pi}\right)^2\tan^2\alpha_{\text{HCS}}},\\
\alpha_{\text{HCS}}&= \arccos\left(\frac{\pi}{2\vartheta_{1/2}}-1\right).
\end{align}
where \(\vartheta_{1/2}\) is a reference value for the waviness of the HCS.
The drift velocities are
\begin{align}
V_{\text{D},r}&=q\frac{2\beta r}{3A}\sqrt{T^2+2Tm}\cot\vartheta
\frac{\Gamma}{(1+\Gamma^2)^2}f,\\
V_{\text{D},\vartheta}&=
q\frac{2\beta r}{3A}\sqrt{T^2+2Tm}
\frac{\Gamma(2+\Gamma^2)}{(1+\Gamma^2)^2}f,\\
V_{\text{HCS},r}&=q\frac{\beta r}{3A}\sqrt{T^2+2Tm}
\frac{\Gamma}{1+\Gamma^2}f'.
\end{align}

\subsubsection{Model \texttt{ref4}}
\label{sec:ref4}

The last two dimensional reference model implemented in \solarprop{} is model \texttt{ref4}~\cite{burger}.
The diffusion tensor agrees with the one of model \texttt{ref3}. The drift effects are modeled in a different way. The regular part agrees with the one from model \texttt{ref2} and is thus given by
\begin{align}
V_{\text{D},r}&=q\frac{2\beta r}{3A}\sqrt{T^2+2Tm}\cot\vartheta
\frac{\Gamma}{(1+\Gamma^2)^2}\left(1-2H\left(\vartheta-\frac{\pi}{2}\right)\right),\\
V_{\text{D},\vartheta}&=
q\frac{2\beta r}{3A}\sqrt{T^2+2Tm}
\frac{\Gamma(2+\Gamma^2)}{(1+\Gamma^2)^2}\left(1-2H\left(\vartheta-\frac{\pi}{2}\right)\right).
\end{align} 
The model simulates a wavy HCS with tilt angle \(\alpha\) by~\cite{burger95}
\begin{equation}
\vartheta_{\Delta}=\frac{2\mathcal{R}V}{|A|\Omega\cos\alpha}=\frac{2\sqrt{T^2+2Tm}V}{|A||Z|\Omega\cos\alpha}
\end{equation}
and
\begin{equation}
V_{\text{HCS},r}=
\begin{cases}
2\sin\vartheta\frac{\Gamma}{\sqrt{1+\Gamma^2}}
\frac{\beta}{6}\frac{\vartheta_{\Delta}\cos\alpha}{\sin (\alpha+\vartheta_{\Delta})}&
, \frac{\pi}{2}-\alpha-\vartheta_{\Delta} < \vartheta < \frac{\pi}{2}+\alpha+\vartheta_{\Delta}\\
0&, \text{ else}.
\end{cases}
\end{equation}

\subsubsection{Model \texttt{standard2D}}
\label{sec:standard}

The model \texttt{standard2D} incorporates a tilt angle dependent wavy HCS and is tuned against recent cosmic ray data. The diffusion tensor is
\begin{align}
\kappa_{\parallel}&=
\begin{cases}
\kappa_0\frac{|A|}{3B}\beta\cdot 0.1 = \kappa_0\beta \frac{1}{30} \frac{r^2}{\sqrt{1+\Gamma^2}}
&, \mathcal{R} < 0.1 \text{ GV}\\
\kappa_0\frac{|A|}{3B}\beta\mathcal{R}=\kappa_{0}\beta\frac{\sqrt{T^2+2Tm}}{3|Z|}
\frac{r^2}{\sqrt{1+\Gamma^2}}
&, \mathcal{R} \geq 0.1 \text{ GV},
\end{cases}\\
\kappa_{\bot}&=0.02\kappa_{\parallel}
\end{align} 
which results in
\begin{align}
\frac{1}{r^2}\frac{\partial r^2\kappa_{rr}}{\partial r}&=
\frac{2}{r}\left(\kappa_{rr}+(\kappa_{\bot}-\kappa_{\parallel})\frac{\Gamma^2}{(1+\Gamma^2)^2}
+\kappa_{rr}\frac{(2+\Gamma^2)}{2(1+\Gamma^2)}
\right),
\\
\frac{1}{r^2\sin\vartheta}\frac{\partial
\sin\vartheta\kappa_{\vartheta\vartheta}}{\partial \vartheta}&=
\frac{1}{1+\Gamma^2}\frac{\kappa_{\bot}}{r^2}\cot\vartheta.
\end{align}
The normalization of the diffusion tensor \(\kappa_0\) is time dependent (see also equation \ref{equa:kappatime}). 
For the drift effects the model is inspired by~\cite{Bobik:2011ig} and thus similar to model \texttt{ref3}. The only difference is that \(\vartheta_{1/2}\) is related to the tilt angle of the current sheet \(\alpha\) by~\cite{burger}
\begin{equation}
\vartheta_{1/2}=\frac{\pi}{2}-\frac{1}{2}\sin\left(\alpha+\frac{2R_L}{r}\right)=
\frac{\pi}{2}-\frac{1}{2}\sin\left(\alpha+\frac{\sqrt{T^2+2Tm}}{|Z|}\frac{2r}{|A|\sqrt{1+\Gamma^2}}\right).
\end{equation} 

\bibliography{mod}
\bibliographystyle{ArXiv}

\end{document}